\documentclass[journal]{IEEEtran}

\usepackage{graphicx,url}

\usepackage[latin1]{inputenc}  
\usepackage[latin1]{inputenc}
\usepackage[compress]{cite}
\usepackage{graphicx}
\usepackage{amsmath}	
\usepackage{algorithmic}
\usepackage[portugues,ruled,lined]{algorithm2e}	
\usepackage[table]{xcolor}
\usepackage{enumerate}
\usepackage{url,subfigure,graphics,rotating,color} 
\usepackage{epstopdf}
\usepackage[english]{babel}
\usepackage{color}

\begin{document}
%
\title{Covert Attacks in Cyber-Physical\\Control Systems}
%
%
%

\author{\IEEEauthorrefmark{1}, \IEEEauthorrefmark{2}, \IEEEauthorrefmark{3}}

\author{
    \IEEEauthorblockN{Alan Oliveira de Sá\IEEEauthorrefmark{1}\IEEEauthorrefmark{2}, Luiz F. Rust da Costa Carmo\IEEEauthorrefmark{1}\IEEEauthorrefmark{3}, Raphael C. S. Machado\IEEEauthorrefmark{3}}\\
    ~\\
    \IEEEauthorblockA{\IEEEauthorrefmark{1}Institute of Mathematics / NCE - Tércio Pacitti Institute,\\
							            Federal University of Rio de Janeiro, 21.941-901, RJ -- Brazil}\\
    \IEEEauthorblockA{\IEEEauthorrefmark{2}Admiral Wandenkolk Instruction Center -- Brazilian Navy,\\
								     Ilha das Enxadas, Baía de Guanabara -- Rio de Janeiro -- RJ -- Brazil \\
								     alan.oliveira.sa@gmail.com}\\
    \IEEEauthorblockA{\IEEEauthorrefmark{2}National Institute of Metrology, Quality and Technology (Inmetro)\\
								    Av. Nossa Senhora das Graças, 50, Xerém, Duque de Caxias, 25.250-020, RJ -- Brazil
								    \\\{lfrust,rcmachado\}@inmetro.gov.br}
}

\markboth{Final version of this paper available at IEEE TRANSACTIONS ON INDUSTRIAL INFORMATICS (\href{http://dx.doi.org/10.1109/TII.2017.2676005}{http://dx.doi.org/10.1109/TII.2017.2676005})}{SKM: My IEEE article}

\IEEEpubid{\begin{minipage}{\textwidth}\ \\[12pt] \begin{center} Copyright (c) 2017 IEEE. Personal use of this material is permitted. However, permission to use this material for any other purposes must be obtained from the IEEE by sending a request to pubs-permissions@ieee.org.
The final version of this paper is available at \href{http://dx.doi.org/10.1109/TII.2017.2676005}{{\color{blue}{http://dx.doi.org/10.1109/TII.2017.2676005}}}. 
\end{center}\end{minipage}}



\maketitle

\begin{abstract}
The advantages of using communication networks to interconnect controllers and physical plants motivate the increasing number of Networked Control Systems, in industrial and critical infrastructure facilities. However, this integration also exposes such control systems to new threats, typical of the cyber domain. In this context, studies have been conduced, aiming to explore vulnerabilities and propose security solutions for cyber-physical systems. In this paper, it is proposed a covert attack for service degradation, which is planned based on the intelligence gathered by another attack, herein proposed, referred as System Identification attack. The simulation results demonstrate that the joint operation of the two attacks is capable to affect, in a covert and accurate way, the physical behavior of a system.
\end{abstract}


%
\IEEEpeerreviewmaketitle

\section{Introduction} \label{sec:Introduction}

\begin{figure}[H]
\begin{center}
\includegraphics[trim=0cm 0.4cm 0cm 0cm, clip=true,width=1\linewidth]{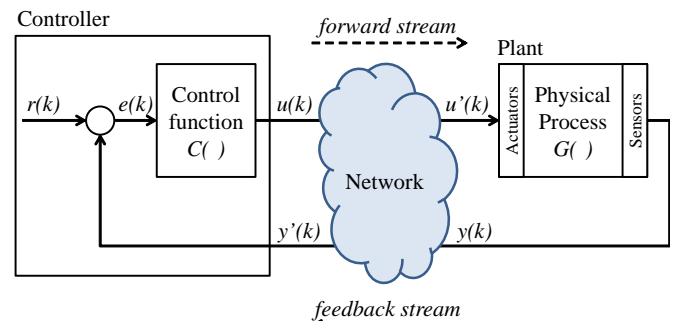}
\caption{Networked Control Systems (NCS)}
\label{fig:NCS}
\end{center}
\end{figure}

\IEEEPARstart{T}{he} integration of the systems used to control physical processes via communication networks aims to assign to such systems better operational and management capabilities, as well as reduce its costs. 
Motivated by these advantages, there is a trend to have an increasing number of industrial process and critical infrastructure systems driven by Networked Control Systems (NCS) \cite{tipsuwan2003implementation,gupta2010networked,zhang2013security,farooqui2014cyber}, also referred as Network-Based Control Systems (NBCS) \cite{chow2001network,long2005denial}. A NCS, shown in Figure \ref{fig:NCS}, consists of a physical plant, described by its transfer function $G(z)$, a controller, which runs a control function $C(z)$, and a communication network that interconnect both devices through a forward stream and a feedback stream. The forward stream connects the output of the controller to the plant's actuators. The feedback stream connects the output of the plant's sensors to the controller's input.

%

At the same time it brings several advantages, the integration of controllers and physical plants in a closed loop through a communication network also exposes such control systems to new threats, typical of the cyber domain. In this context, studies have been conduced aiming to characterize vulnerabilities and propose security solutions for the NCSs.
\vspace{-4mm}

One possible way to attack a NCS is by hacking its software, {\it i.e.} changing the configuration or even the code executed by the controller, following a strategy similar to that used by the Stuxnet worm \cite{langner2011stuxnet}. Another possible way for an attacker to negatively affect a NCS is by interfering on its communication process. Basically, an attacker may interfere in the forward and/or the feedback streams by three different means: inducing a jitter, causing data loss due to packet drop outs, or even injecting false data in the communication process.

In this work, it is developed an attack where false data is injected in the communication process of a NCS. It is demonstrated the possibility to degrade the service performed by a plant, through interventions that produce subtle changes on its physical behavior. Such interventions aim to reduce the efficiency of the plant or even cause it damage in mid/long therm. It is worth mentioning that an uncontrolled intervention in a NCS may lead the plant to an immediate breakdown, or even significantly change its behavior, which may cause the disclosure of the attack and the eventual failure of the operation. Thus, the changes driven by the attack herein proposed are dimensioned so that the modifications in the plant's behavior are physically difficult to be perceived. That is why the present attack is classified as physically covert.

To ensure that the attack to a NCS is physically covert, the attacker must plan his offensive based on an accurate knowledge about the system dynamics, otherwise the consequences of the attack may be unpredictable. 
One possible way to obtain such knowledge is through conventional intelligence operations, performed to collect information regarding to the design and the dynamics of the NCS. 
Another way to gather information about the targeted system is through what we refer in this paper as a {\it Cyber-Physical Intelligence} attack. 
To this end, it is also proposed in this paper the {\it System Identification} attack, which aims to collect information about the plant's transfer function $G(z)$ and the controller's control function $C(z)$ of a NCS. This attack is proposed based on the Backtracking Search Optimization algorithm (BSA) \cite{civicioglu2013backtracking}. 

This work motivated the formalization of a number of concepts related to covertness and intelligence in the context of the cyber-physical security. Thus, an additional contribution of this paper is the proposition of a terminology that encompasses a whole new class of attacks on cyber-physical systems. The proposed taxonomy establishes a new approach regarding to the covertness of attacks on cyber-physical systems, which must be analyzed from two aspects simultaneously: the physical and the cybernetic aspects.

It is worth mentioning that the objective of this work is not to facilitate covert attacks for service degradation in cyber-physical control systems. The purpose of this work is to demonstrate the degree of accuracy that may be achieved in this kind of attack, especially when supported by System Identification attacks and, therefore, encourage the research for countermeasures to such attacks. 
The rest of this paper is organized as follows: First, in Section \ref{sec:trab_rel}, some related works are presented. Later, in Section \ref{sec:taxonomy}, it is proposed a taxonomy regarding to the cyber-physical attacks that may happen in the control loop of a NCS. In Section \ref{sec:algoritmo}, it is described a System Identification attack. Then, in Section \ref{sec:ataque_furtivo}, it is defined a cover attack for service degradation. After that, in Section~\ref{Resultados}, there are reported the results obtained by simulations of covert attacks for service degradation, supported by System Identification attacks. Finally, in Section \ref{Conclusion}, some conclusions and possible future works are presented.

\section{Related Works}\label{sec:trab_rel}

The possibility of cyber-physical attacks became a reality after the launch of the Stuxnet worm \cite{langner2011stuxnet} and has been motivating researches concerning the security of NCSs. In this section, some works related to this subject are presented.

In \cite{long2005denial} the authors propose two queueing models to evaluate the impact of delay jitter and packet loss in a NCS under attack. The attack is not designed taking into account a previous knowledge about the models of the controller and the physical plant. Thus, to affect the plant's behavior, the attacker arbitrarily floods the network with an additional traffic, causing jitter and packet loss. In this tactics, the excess of packets in the network can reduce the covertness of the attack, allowing the adoption of countermeasures, such as packet filtering \cite{long2005denial} or blocking the malicious traffic on its origin \cite{snoeren2002single}. Additionally, the arbitrary intervention in a system which the model is unknown may lead the plant to an extreme physical behavior, which is not desired if it is intended a covert attack.

In \cite{farooqui2014cyber}, the authors present a Supervisory Control and Data Acquisition (SCADA) testbed using TrueTime, a MATLAB/Simulink based tool. They demonstrate an attack where a malicious agent sends false signals to the controller and to the actuator of a NCS. In that paper, the false signals are randomly generated aiming to make a DC motor lose its stability. This kind of attack does not require a previous knowledge about the plant and controller of the NCS. On the other hand, the desired physical effect and the covertness of the attack can not be ensured due to the unpredictable consequences of the application of random false signals to a system which the model is unknown.

More recently, in \cite{teixeira2015secure}, the authors give a general framework for the analysis of a wide variety of methods of attack in NCSs. In their classification, it is stated that covert attacks in NCSs require high level of knowledge about the targeted system. Exemples of covert attacks are provided in \cite{Smth2011,smith2015covert}. In these works the attacks are reformed by a man-in-the-middle (MitM), where the attacker needs to know the model of the plant under attack and also inject false data in both the forward and the feedback streams. The covertness of the attacks described in \cite{Smth2011,smith2015covert}, which depends on the difference between the actual model of the plant and the model used by the attacker, is analyzed from the perspective of the signals arriving to the controller, without addressing if the physical effects on the plant are noticeable, or if they are covet when faced by a human observer.

In \cite{teixeira2015secure,Smth2011,smith2015covert}, where it is required a previous knowledge about the models of the NCS under attack, it is not described how the knowledge about the system is obtained by the attacker. It is just stated that a model is previously known to subsidize the design of the attack. The joint operation, herein proposed, of a covert attack for service degradation, supported by a System Identification attack, aims to fill this hiatus, demonstrating how the data of a NCS can be obtained and how a covert attack can take advantage from it.
The Table~\ref{tab:resumo_rel_work} presents a synthesis of the characteristics of the attacks referred in this section.
\vspace{-5mm}
\begin{table}[H]
\centering
\caption{Synthesis of the related attacks}
\label{tab:resumo_rel_work}
\scriptsize{
\begin{tabular}{l|c|c|c}
\hline
 													& 	 						&  System 		&  How the knowledge			\\
Attack 												&  Method						&  knowledge 		&  is obtained			\\
\hline\hline 
 Stuxnet {\it worm} \cite{langner2011stuxnet} 					& Modifications in 			 	&  Yes			& Experiments in		\\
  													& the PLC code					&  				& a real system			\\
\hline
 Long, {\it et al.} \cite{long2005denial} 						& Inducing {\it jitter} 				&  None			& N/A				\\
  													& and packet loss 		 		&  				& 					\\
 \hline
 Farooqui, {\it et al.} \cite{farooqui2014cyber}					& Data injection  				&  None			& N/A 				\\
 \hline
 Smith \cite{Smth2011,smith2015covert}						& Data injection  				&  Yes			& Not described			\\
\hline
\end{tabular}
}
\end{table} 

\section{Taxonomy}\label{sec:taxonomy}

In this section it is presented a taxonomy concerning the possible attacks on cyber-physical control systems.
In Section \ref{subsec:classificacao_ataques}, the attacks are briefly described and classified according to the way they act in the NCS. In Section \ref{subsec:stealthiness}, it is proposed a new approach for the analysis of the covertness of attacks in cyber-physichal systems.

\subsection{Classification of the Attacks}\label{subsec:classificacao_ataques}

The attacks to cyber-physichal control systems may take place on its devices -- {\it i.e.} the controller, and the plant's sensors and actuators -- and/or on its communication system, affecting the forward and the feedback streams. As a premise, we must consider that the  {\it service} intended to be attacked/protected in such system is the work performed by the physical process, controlled by the NCS.

\begin{figure}[ht]
\begin{center}
\includegraphics[trim=0cm 0cm 0cm 0cm, clip=true,width=1\linewidth]{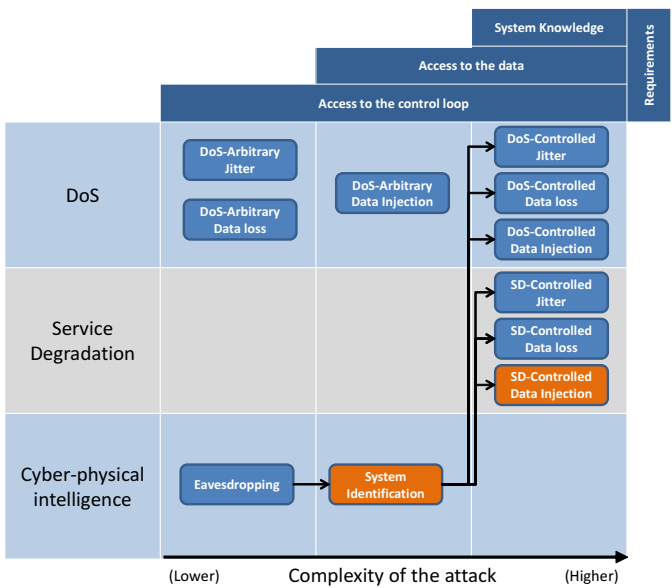}
\caption{Classification and requirements of the cyber-physical attacks that act in the control loop of a NCS.}
\label{fig:quadro_ataques}
\end{center}
\end{figure}

 Considering the aforementioned definition of service in a NCS, the attacks may be classified within three different categories, as shown in Figure \ref{fig:quadro_ataques}:
\begin{itemize}
	\item Denial-of-Service (DoS) \cite{hussain2003framework}: in a NCS, the DoS attacks comprehends all kind of cyber-physical attacks that deny the operation of the physical process, interrupting the execution of the work, or service, that the controlled plant is intended to do. The attack results, for example, in behaviors that may shut the plant down or even destroy it in a short therm.
	\item Service Degradation (SD): the SD attacks consist of malicious interventions that are done in the control loop in order to reduce the efficiency of the service, {\it i.e.} the efficiency of the physical process, or even reduce the mean time between failure (MTBF) of the plant in mid therm or long therm.
	\item Cyber-physical Intelligence (CPI): the concept of Cyber-physical Intelligence, herein proposed, is different from the concept of where cyber-physical systems are integrated with intelligent systems \cite{ramos2011cyber}. In the present taxonomy, the CPI attacks comprehend actions that are performed in the control loop of a NCS in order to gather information about the system's operation and/or its design. This attacks are intended to acquire the intelligence necessary to plan covert and controlled attacks, or even to subsidize data for replay attacks \cite{langner2011stuxnet}.
\end{itemize}

In Figure \ref{fig:quadro_ataques}, six kinds of DoS attacks are presented, with their respective requirements. From this six DoS attacks, the less complex are the arbitrary ones:
\begin{itemize}
	\item DoS-Arbitrary Jitter: in this kind of attack, the delay of the forward and/or the feedback stream is arbitrarily changed, without a previous knowledge about the models of the NCS, in order to lead the system to an instability or to a condition that causes the interruption of the physical process. This attack only requires access to the control loop, once it may be performed by just consuming the resources of the system, such as the bandwidth of communication links, or the computational resources of the equipments that are in the control loop.
	\item  DoS-Arbitrary Data Loss: in this kind of attack, the attacker prevents the data from reaching the actuator and/or the controller of the NCS. The communication channel is jammed arbitrarily, without a previous knowledge about the models of the NCS, leading the system to an instability or to a condition that causes the interruption of the physical process. It is worth mentioning that some DoS-Arbitrary Jitter attack may result in a DoS-Arbitrary Data Loss attack, if an eventual higher delay cause packet drop outs. As the DoS-Arbitrary Jitter attack, this attack only requires access to the control loop of the NCS.
	\item DoS-Arbitrary Data Injection: in such attacks, the attacker sends arbitrary false data to the controller, as it was sent by the sensors, and/or to the actuators, as it was sent by the controller. The false data is injected in the NCS closed loop without a previous knowledge about the models of the NCS. This attack is more complex than the DoS-Arbitrary Jitter and the DoS-Arbitrary Data Loss attacks, given that it requires access to the data that flows in the control loop of the NCS.
\end{itemize}

The attacks classified as DoS-controlled -- DoS-Controlled Jitter, DoS-Controlled Data Loss, and DoS-Controlled Data Injection -- shown in Figure \ref{fig:quadro_ataques} interfere in the control loop of a NCS by the same means that their respective DoS-Arbitrary attacks. The difference between a DoS-Controlled attack and a DoS-Arbitrary attack is that, in the former, the interference caused by the attacker is precisely planned and executed, in order to achieve the exact desired behavior that leads the physical service to an interruption, in a more efficient way. Thus, to achieve such efficiency, a DoS-Controlled attack require an accurate knowledge about the NCS models, {\it i.e.} the plant and the controller transfer functions, which have to be analyzed to plan the attack. 

Regarding to the SD attacks, we must consider the three different kinds of attack shown in Figure \ref{fig:quadro_ataques}: SD-Controlled Jitter, SD-Controlled Data Loss, and SD-Controlled Data Injection. The difference between a SD-Controlled attack and a DoS-Controlled attack is that the former is not intended to interrupt the physical process in a short therm. It aims to keep the process running with reduced efficiency, sometimes also targeting a gradual physical deterioration of the controlled devices. 
To succeed, the SD-Controlled attacks need to be planned based on an accurate knowledge about the dynamics and the design of the NCS. Otherwise, the attack can eventually interrupt the physical process, due to unpredicted reasons, evolving to a DoS attack. 

The system knowledge required to both DoS-Controlled and SD-Controlled attacks, can be gathered through CPI attacks, as shown in Figure \ref{fig:quadro_ataques}. The first, and simpler, CPI attack is the eavesdropping attack \cite{khatri2015taxonomy,zou2016intercept}, which consists of simply capturing the data transmitted through the forward and feedback streams of the NCS. The second CPI attack, herein proposed, is the System Identification attack, which aims to obtain information about the control function of the controller and the transfer function of the plant, by analyzing the signals that flow in the network between the controller and the plant. The CPI attacks by themselves do not impact on the NCS, but they are an useful tool to plan efficient and accurate DoS-Controlled and SD-Controlled attacks.

\subsection{Cybernetic vs. Physical Covertness}\label{subsec:stealthiness}

The covertness of an attack regards to its capacity to not be perceived or detected. In the case of cyber-physical attacks on NCSs, the covertness must be simultaneously analyzed in two different domains: the cyber domain; and the physical domain. In this sense, it is presented in this section the definition of what is a {\it cybernetically covert} attack and what is a {\it physically covert} attack:
\begin{itemize}
	\item Cybernetically covert attacks: are the attacks that have low probability to be detected by algorithms that monitor the softwares, communications and data of the NCS, or by systems that monitor the dynamics of the plant.
	\item Physically covert attacks: are attacks that cause physical effects that can not be easily noticed or identified by a human observer. The attack slightly modifies some behaviors of the system in a way that it physically affects the plant, but the effect is not easily perceptible or it can eventually be understood as a consequence of another root cause, other than an attack.
\end{itemize}


\section{The System Identification Attack}\label{sec:algoritmo}

The System Identification attack, herein proposed, is intended to assess the coefficients of the plant's transfer function $G(z)$ and the controller's control function $C(z)$. Both functions are linear time-invariant (LTI). The attack uses the BSA metaheuristic, proposed in \cite{civicioglu2013backtracking} and briefly described in \cite{de2016distributed}, to minimize the fitness function presented in this section.

The BSA is an evolutionary algorithm that uses the information obtained by past generations -- or iterations -- to perform the search for solutions for optimization problems. The algorithm has two parameters that are empirically adjusted: the size of its population $P$; and $\eta$, described in \cite{de2016distributed}, that establishes the amplitude of the movements of the individuals of $P$. The parameter $\eta$ must be adjusted aiming to assign to the algorithm both good exploration and exploitation capabilities.

If the input $i(k)$ and the output $o(k)$ of a device of the NCS is known, the model of such device can be assessed by applying the known $i(k)$ in an estimated model, which must be adjusted until its estimated output $\hat{o}(k)$ converge to $o(k)$. In this sense, the BSA is used to iteratively adjust the estimated model, by minimizing a specific fitness function, until the estimated model converge to the actual model of the real device, that can be a controller or a plant of the NCS. 

To establish the fitness function, firstly, it must be considered a generic LTI system, whose transfer function $Q(z)$ is represented by (\ref{equ:z_tf}):
\begin{equation}
Q(z)=\frac{O(z)}{I(z)} = \frac{a_nz^{n}+a_{n-1}z^{n-1}+...+a_{1}z^{1}+a_{0}}{z^{m}+b_{m-1}z^{m-1}+...+b_{1}z^{1}+b_{0}},
\label{equ:z_tf}
\end{equation}
wherein $I(z)$ is the input of the system, $O(z)$ is the output of the system, $n$ and $m$ are the order of the numerator and the denominator, respectively, and $[a_n,a_{n-1},...a_1, a_0]$ and $[b_{m-1},b_{m-2},...b_1, b_0]$ are the coefficients of the numerator and the denominator, respectively, that are intended to be found by this System Identification attack.
Also, it must be considered that $i(k)$ and $o(k)$ represent the sampled input and output of the system, respectively, where $I(z)=\mathcal{Z}[i(k)]$, $O(z)=\mathcal{Z}[o(k)]$, $k$ is the number of the sample and $\mathcal{Z}$ represents the Z-transform operation.  

In this System Identification attack, $i(k)$ and $o(k)$ are firstly captured by an eavesdropping attack \cite{khatri2015taxonomy,zou2016intercept}, for exemple, during a monitoring period $T$. To deal with the eventual loss of samples, that may not be received by the attacker during $T$, the algorithm holds the value of the last received sample, according with (\ref{equ:hold_input}), wherein $x(k)$ can either be $i(k)$ or $o(k)$:
\begin{equation}
x(k)=\left\{\begin{array}{ll}
	x(k-1)&\;\;\;\; \mbox{if the sample $k$ is lost;}\\
&\\
	x(k)&\;\;\;\; \mbox{otherwise}.\\
\end{array} \right.
\label{equ:hold_input}
\end{equation}

Then, after acquiring $i(k)$ and $o(k)$, the captured $i(k)$ is applied to the input of an estimated model, that is described by a transfer function whose coefficients $[a_{n,j},a_{n-1,j},...a_{1,j},a_{0,j},b_{m-1,j}$, $b_{m-2,j},...b_{1,j},b_{0,j}]$ are the coordinates of an individual $j$ of the BSA. The application of $i(k)$ to the input of the estimated model results in an output signal $\hat{o}_j(k)$. After obtaining $\hat{o}_j(k)$, the fitness $f_j$ of the individual $j$ is computed comparing the output $o(k)$, captured from the attacked device, with the output $\hat{o}_j(k)$ of the estimated model, according with (\ref{equ:fitness}):
\begin{equation}
f_j = \frac{\sum\limits_{k=0}^N(o(k)-\hat{o}_j(k))^2}{N},
\label{equ:fitness}
\end{equation}
wherein $N$ is the number of samples that exist during the monitoring period $T$. Note that, if the attacker do not lose any sample of $i(k)$ and $o(k)$ during $T$, then $\min{f_j}=0$ when $[a_{n,j},a_{n-1,j},...a_{1,j},a_{0,j},b_{m-1,j},b_{m-2,j},...b_{1,j},b_{0,j}]=[a_n,a_{n-1},...a_1, a_0,b_{m-1},b_{m-2},...b_1, b_0]$, {\it i.e.} when the estimated model converges to the actual model of the attacked device.

It is possible to establish an analogy between this System Identification attack and the Known Plaintext cryptanalytic attack \cite{stallings2006cryptography}, wherein $i(k)$ and $o(k)$ correspond to the plaintext and ciphertext, respectively, the form of the generic transfer function $Q(z)$ corresponds to the encryption algorithm and the actual coefficients of $Q(z)$ corresponds to the secret key.

\section{The Covert Attack for Service Degradation}\label{sec:ataque_furtivo}

Based on the taxonomy presented in Section \ref{subsec:classificacao_ataques}, the attack described in this section is classified as a SD-Controlled Data Injection attack. Its purpose is to reduce the MTBF of the plant and/or reduce the efficiency of the physical process that the plant performs, by inserting false data in the control loop. At the same time, the attacker desires that the attack meets the requirement of being physically covert, as the definition presented in Section~\ref{subsec:stealthiness}.

One way to degrade a physical service is through the induction of an overshoot during the transient response of a plant. The overshoots, or peaks occurred when the system exceeds the targeted value during the transient response, can cause stress and possibly damage physical systems such as mechanical, chemical and electromechanical systems \cite{el1989variable,tran2007robust}. Additionally, once they occur in a short period of time, the overshoots are difficult to be noticed by a human observer. Another way to degrade the service of a plant is causing a constant steady state error on it, {\it i.e.} producing a constant error when $t\rightarrow \infty$. A low proportion steady state error, besides being difficult to be perceived by a human observer, may reduce the efficiency of the physical process or, occasionally, stress and damage the system in mid/long therm.

\begin{figure}[H]
\begin{center}
\includegraphics[trim=0cm 0cm 0cm 0cm, clip=true,width=1\linewidth]{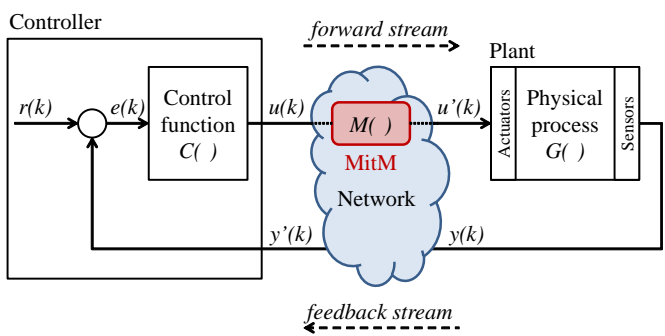}
\vspace{-6mm}
\caption{MitM attack}
\label{fig:MitM}
\end{center}
\end{figure}

In the present attack, to achieve either of the two mentioned effects, {\it i.e.} an overshoot or a constant steady state error, the attacker interfere in the NCS's communication process by injecting false data into the system in a controlled way. To do so, the attacker act as a MitM that executes an attack function $M(z)$, as presented in Figure~\ref{fig:MitM}, wherein $U'(z)=M(z)U(z)$, $U(z)=\mathcal{Z}[u(k)]$ and $U'(z)=\mathcal{Z}[u'(k)]$. The function $M(z)$ is designed based on the models of the plant and the controller, both obtained through the System Identification attack, described in Section \ref{sec:algoritmo}. The effectiveness of the attack, therefore, depends on the design of $M(z) $, which in turn depends on the accuracy of the System Identification attack. It is worth mentioning that, in Figure \ref{fig:MitM}, although the MitM is placed in the forward stream, it is also possible perform an attack by interfering in the feedback stream of the NCS. The MitM may act in wired or wireless networks, such as in \cite{hwang2008study}.
\vspace{-4mm}
\vspace{-4mm}

\section{Results}\label{Resultados}

In this section, there are presented the results obtained through simulations that combines the System Identification attack with physically covert SD-Controlled attacks. First, in Section \ref{subsec:modelo_sistema}, the model of the attacked system is described. Then, in Section \ref{subsec:res_ident}, there are presented the results obtained by the System Identification attack. After that, in Section \ref{subsec:res_SD}, there are presented the results achieved by simulations of physically covert SD-Controlled Data Injection attacks, planned based on the data gathered by the System Identification attack.

\subsection{The Model of the Attacked System}\label{subsec:modelo_sistema}

The attacked NCS has the same architecture of the NCS shown in Figure \ref{fig:NCS}, and consists of a Proportional-Integral (PI) controller that controls the rotational speed of a DC motor. The PI control function $C(z)$ and the DC motor transfer function $G(z)$ are the same as in \cite{long2005denial}. The equations of this NCS are represented in (\ref{equ:c_g_z}):
\begin{equation}
		C(z)= \frac{c_1z-c_2}{z-1} \qquad \qquad G(z)= \frac{g_1z+g_2}{z^2-g_3z+g_4}\\
\label{equ:c_g_z}
\end{equation}
wherein $c_1=0,1701$, $c_2=-0,1673$, $g_1=0,3379$, $g_2=0,2793$, $g_3=-1,5462$ and $g_4=0,5646$.  The sample rate of the system is 50 samples/s and the set point $r(k)$ is an unitary step function. The network delay is not taken into account in the simulations of this paper.

\subsection{Results of the System Identification Attack}\label{subsec:res_ident}

In this Section, the performance of the System Identification attack is evaluated through a set of simulations performed in MATLAB. The SIMULINK tool is used to compute the output  $\hat{o}_j$ of the estimated models, whose coefficients are the coordinates of an individual $j$ of the BSA.

The structure of the equations represented in (\ref{equ:c_g_z}) are previously known by the attacker once that, as a premise, it is known that the target is a NCS that controls a DC motor using a PI controller. In these simulations, the goal of the System Identification attack is to discover $g_1$, $g_2$, $g_3$, $g_4$, $c_1$ and $c_2$, also taking into account scenarios in which the attacker occasionally loses samples of the forward and feedback streams.

Each time that the DC motor is turned on, the forward and the feedback streams are captured by the attacker during a period $T=2s$. All initial conditions are considered 0, by the time that the motor is turned on. The coefficients of $G(z)$, $[g_1,g_2,g_3,g_4]$, and the coefficients of $C(z)$, $[c_1,c_2]$, are computed separately considering that, albeit the closed loop, $G(z)$ and $C(z)$ are independent transfer functions. 
To assess $[g_1,g_2,g_3,g_4]$, the attacker considers the forward stream as the input and the feedback stream as the output of the estimated plant. In the opposite way, to assess $[c_1,c_2]$, the attacker considers the feedback stream as the input and the forward stream as the output of the estimated controller.

To simulate the loss of samples, it is considered four different rates $l$ of sample loss: 0, 0.05, 0.1 and 0.2. Thus, a sample is lost by the attacker every time that $l < \mathcal{P}$, where $\mathcal{P}\sim U(0,1)$ and $U$ is the uniform distribution. There are executed 100 different simulations for each rate of sample~loss.

In the BSA, the population is set to 100 individuals and $\eta$, empirically adjusted, is 1. To assess the coefficients of the controller $[c_1,c_2]$, the algorithm is executed for 600 iterations. To assess the coefficients of the plant $[g_1,g_2,g_3,g_4]$, the number of iterations is increased to 800, due to the higher number of dimensions of the search space in this case. The limits of each dimension of the search space are $[-10,10]$.

\begin{figure*}[!ht]
\begin{center}
\begin{tabular}{ccc}
\subfigure[$g_1$ of $G(z)$]{
\includegraphics[trim=0cm 0cm 0cm 0cm, clip=true,width=.3\linewidth]{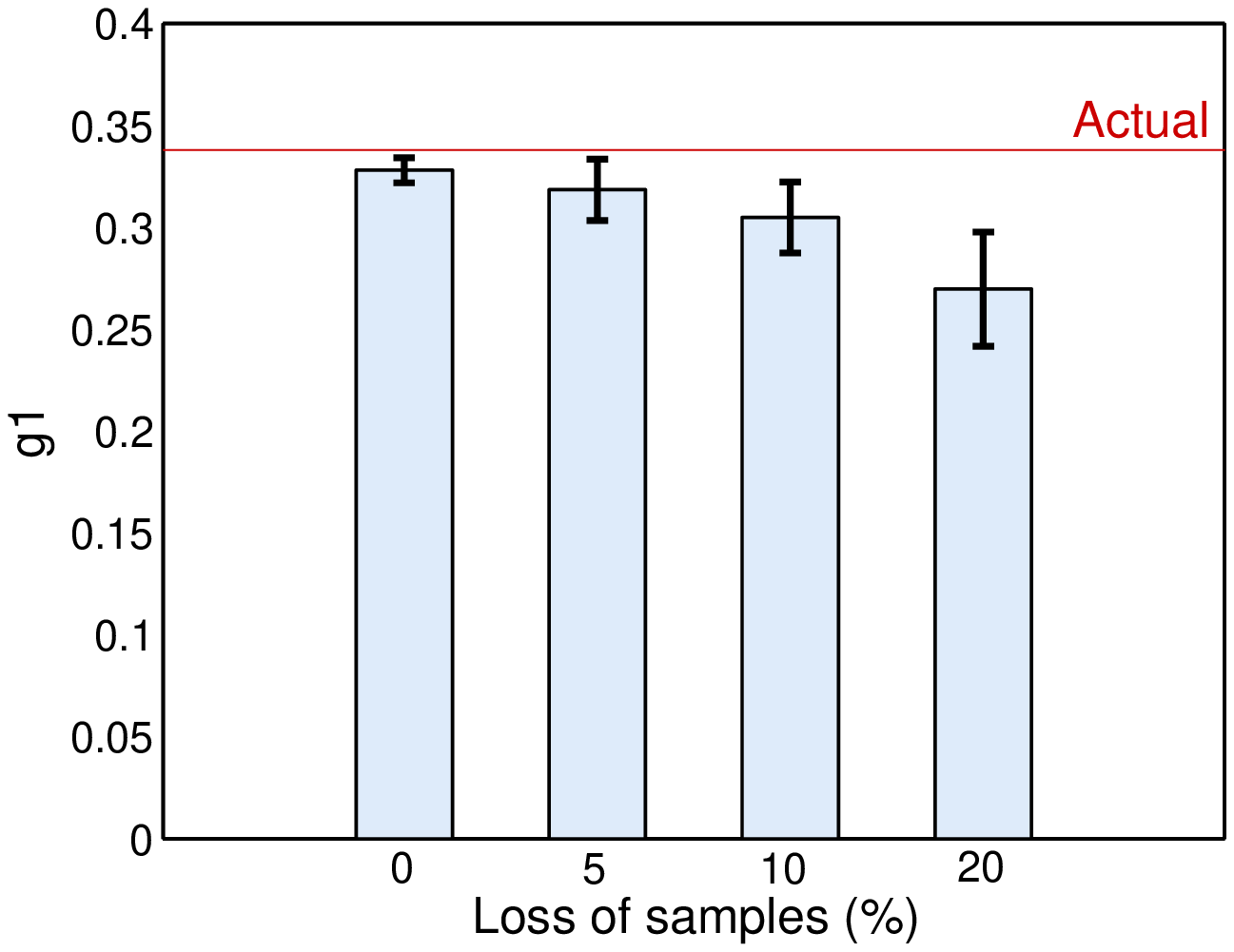}
\label{fig:media_ic_g1}
}&
\subfigure[$g_2$ of $G(z)$]{
\includegraphics[trim=0cm 0cm 0cm 0cm, clip=true,width=.3\linewidth]{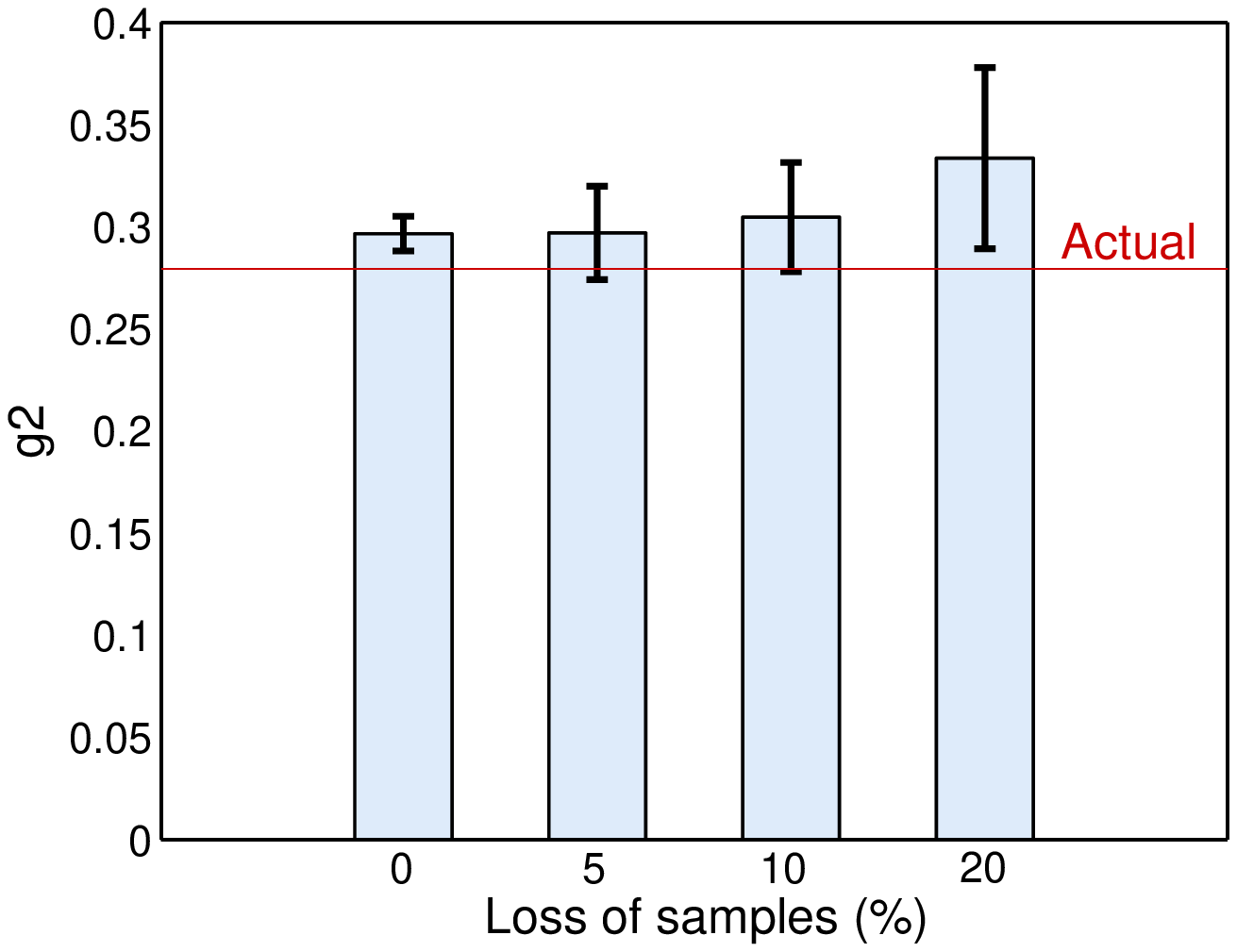}
\label{fig:media_ic_g2}
}&
\subfigure[$g_3$ of $G(z)$]{
\includegraphics[trim=0cm 0cm 0cm 0cm, clip=true,width=.3\linewidth]{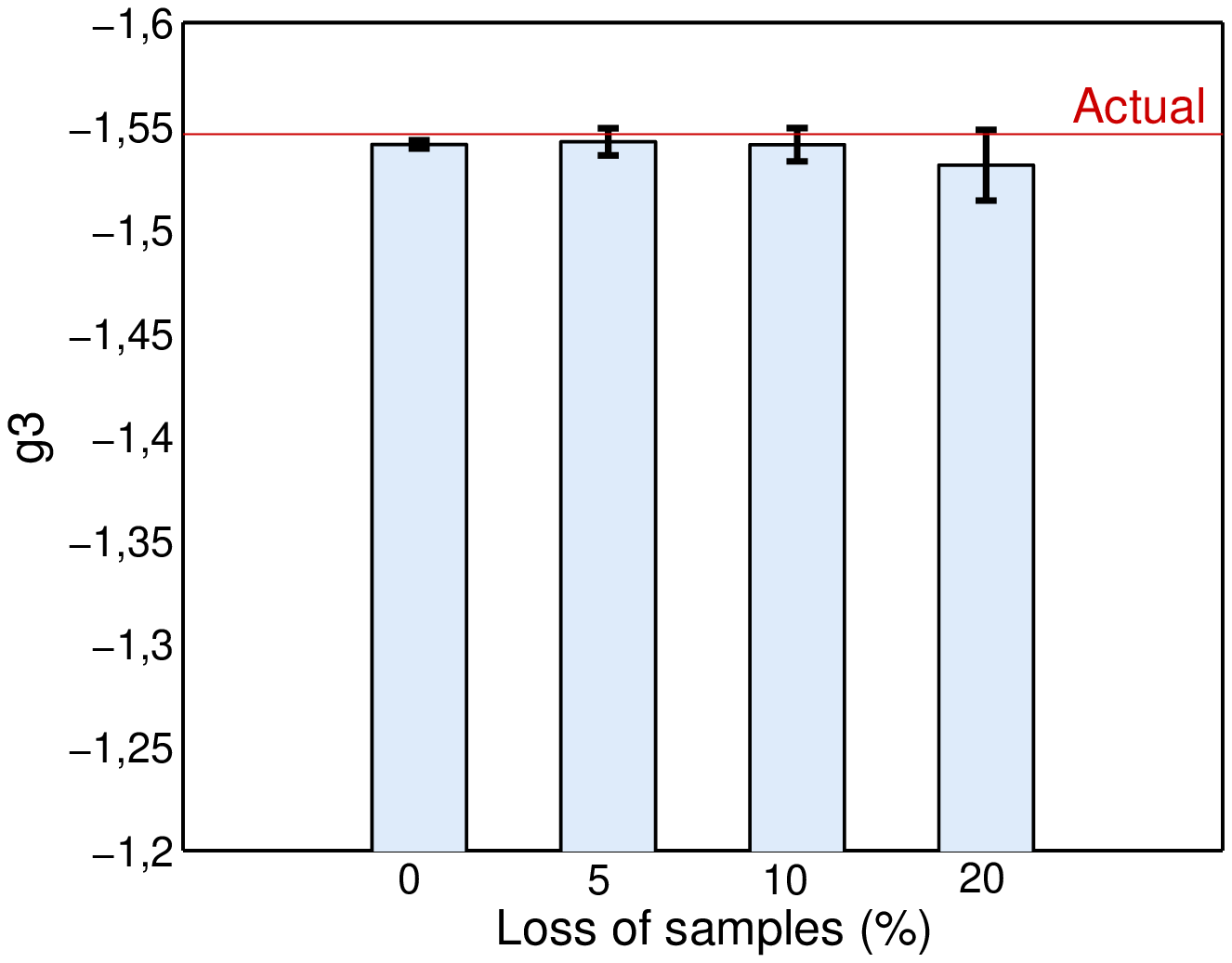}
\label{fig:media_ic_g3}
}\\
\subfigure[$g_4$ of $G(z)$]{
\includegraphics[trim=0cm 0cm 0cm 0cm, clip=true,width=.3\linewidth]{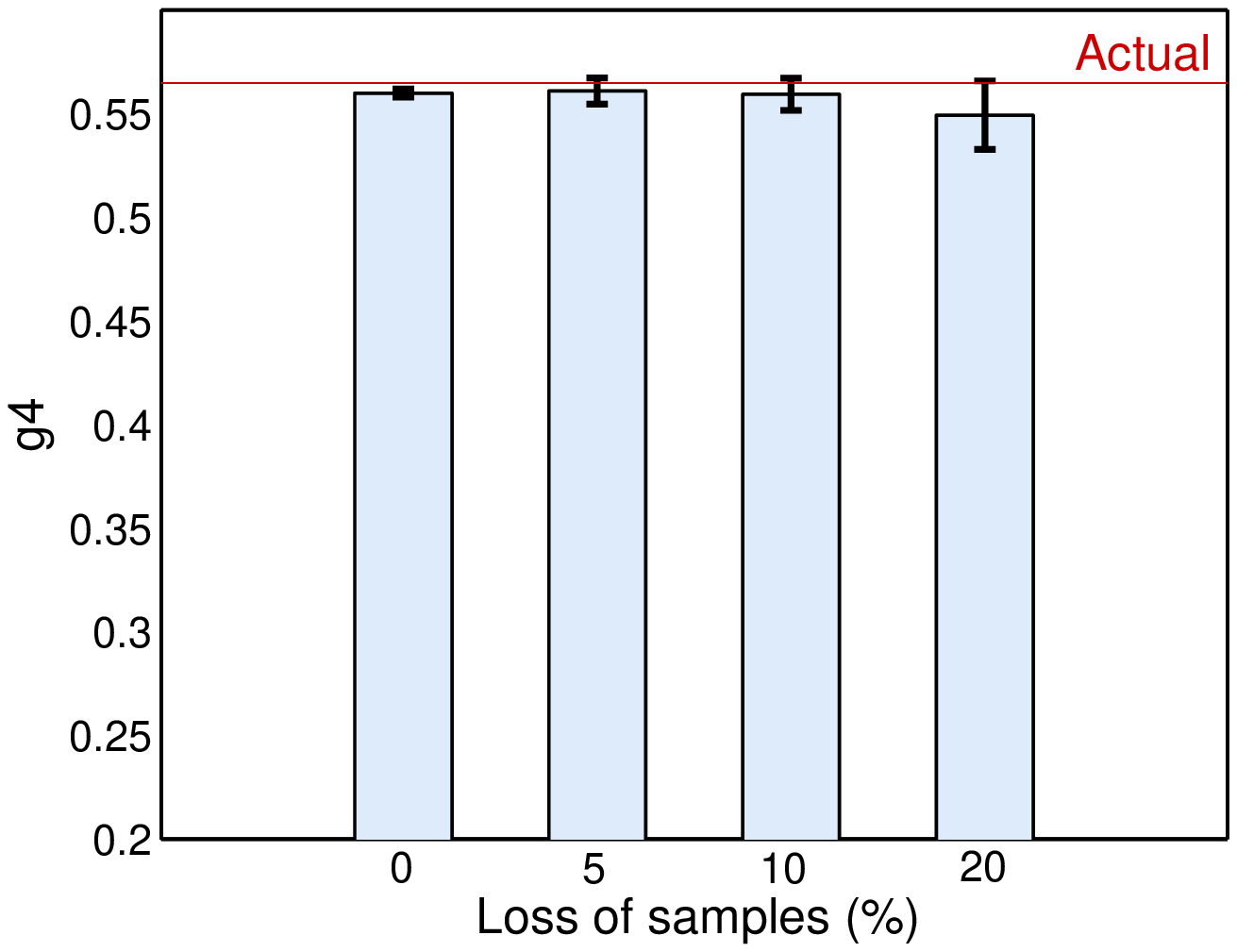}
\label{fig:media_ic_g4}
}&
\subfigure[$c_1$ of $C(z)$]{
\includegraphics[trim=0cm 0cm 0cm 0cm, clip=true,width=.3\linewidth]{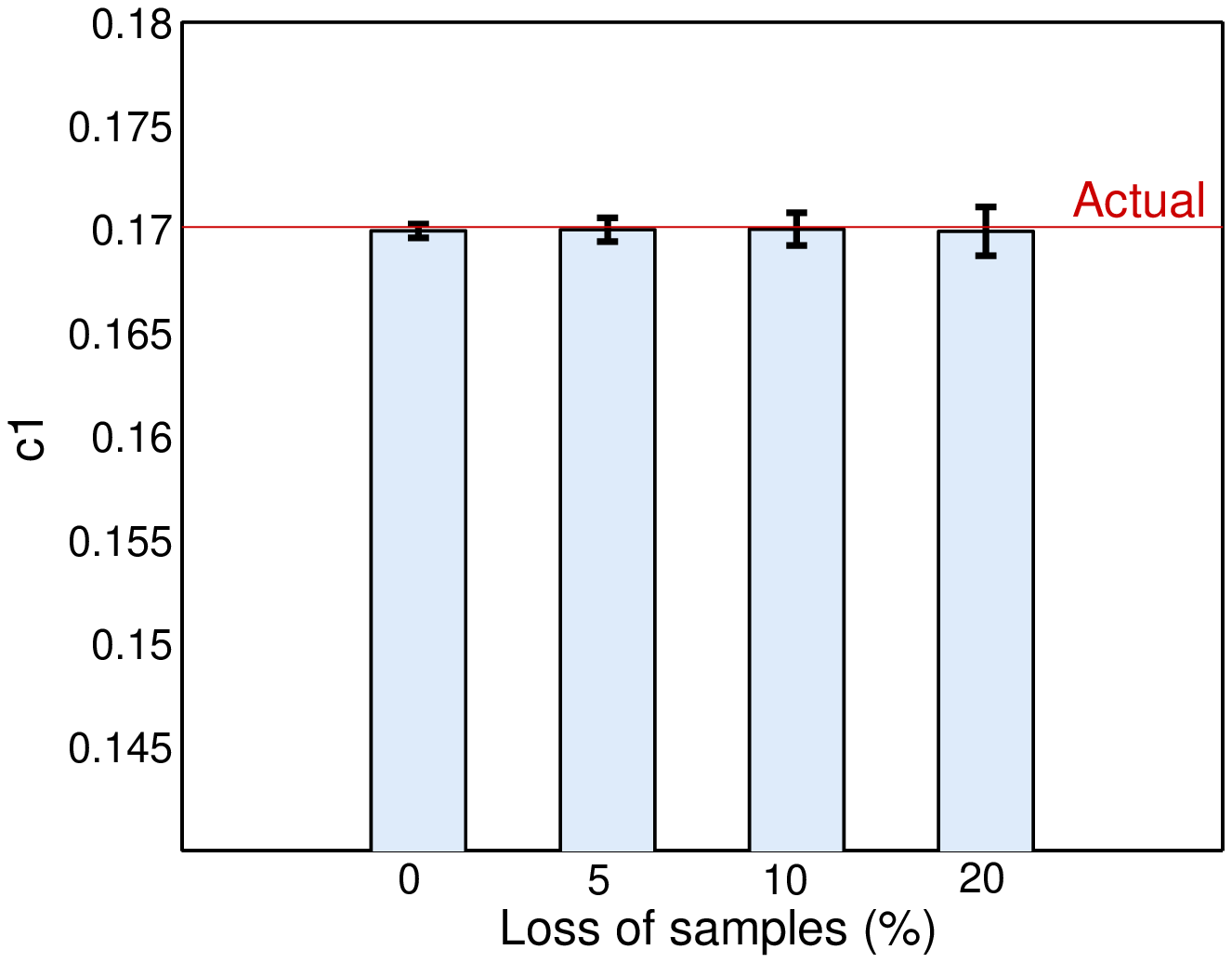}
\label{fig:media_ic_c1}
}&
\subfigure[$c_2$ of $C(z)$]{
\includegraphics[trim=0cm 0cm 0cm 0cm, clip=true,width=.3\linewidth]{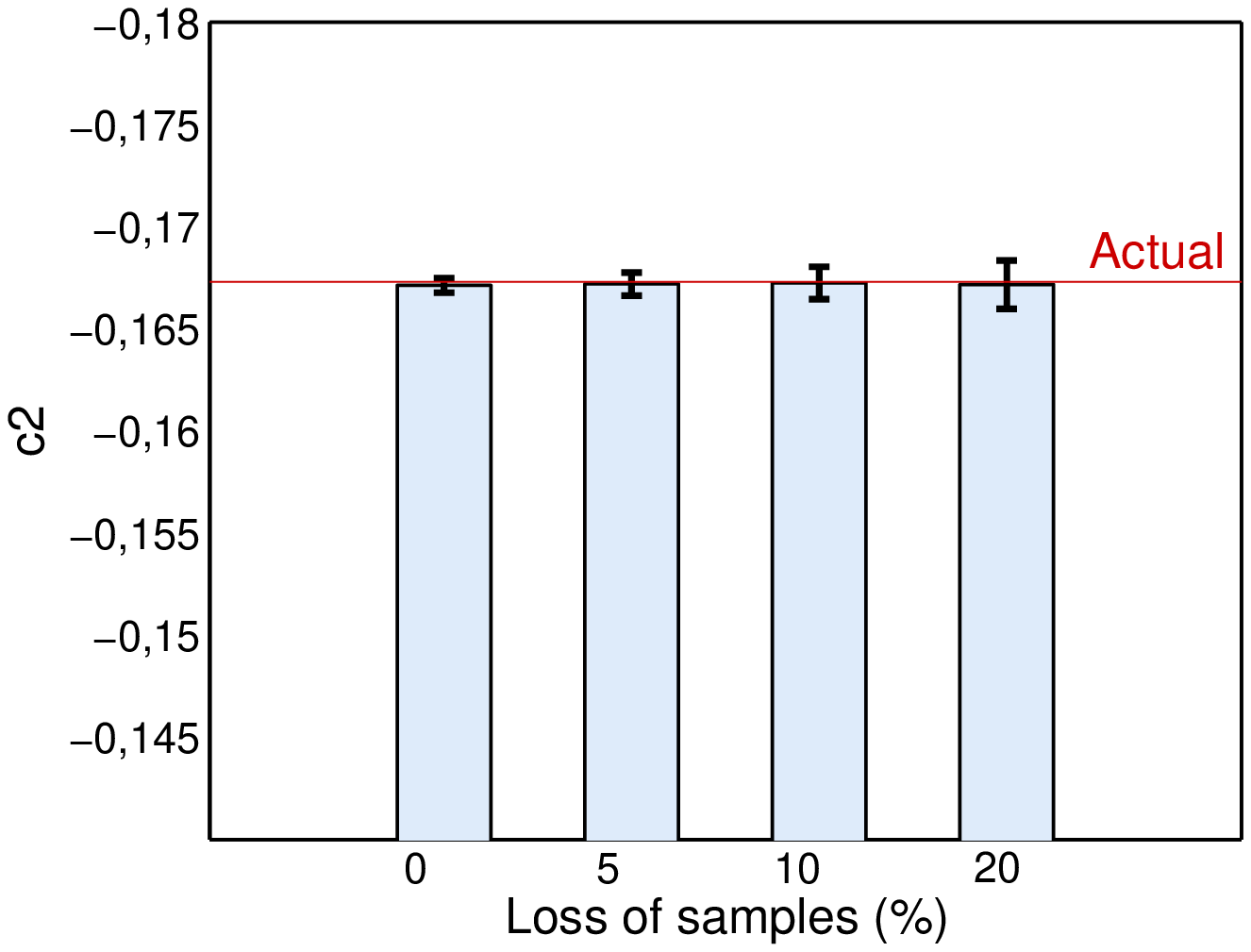}
\label{fig:media_ic_c2}
}\\
\end{tabular}
\caption{Mean, with a CI of 95\%, of the estimated coefficients of $G(z)$ and $C(z)$, in face of different rates of sample loss.}
\label{fig:mean_ic}
\end{center}
\end{figure*}

Figure \ref{fig:mean_ic} shows the means of 100 estimated values of $g_1$, $g_2$, $g_3$, $g_4$, $c_1$ and $c_2$, with a Confidence Interval (CI) of 95\%, considering different rates of sample loss. The actual values of the coefficients of $C(z)$ and $G(z)$ are also depicted in Figure~\ref{fig:mean_ic}. Note that the scales of Figures \ref{fig:media_ic_g1}, \ref{fig:media_ic_g2}, \ref{fig:media_ic_g3} and \ref{fig:media_ic_g4} are different from the scales of Figures \ref{fig:media_ic_c2} and \ref{fig:media_ic_c2}, due to the smaller amplitude of the CI of $c_1$ and $c_2$. In Addition, some statistics of the obtained results are presented in Table~\ref{tab:estatisticas}.
\begin{table*}[!ht]
\centering
\caption{Statistics of the results obtained with different rates of sample loss}
\label{tab:estatisticas}
\scriptsize{
\begin{tabular}{l|c|c|c|c|c|c|c|c|c|c|c|c|c|c|c|c|c|c|c|c|c|c|c|c}
\hline
Loss of					&  \multicolumn{6}{c|}{Mean} &  \multicolumn{6}{c}{Standard deviation}		\\
\cline{2-13}	
samples	&$g_1$&$g_2$&$g_3$&$g_4$&$c_1$&$c_2$&$g_1$&$g_2$&$g_3$&$g_4$&$c_1$&$c_2$\\
\hline\hline 
0\%	&	0.32793	&	0.29652	&	-1.54121	&	0.55983	&	0.16991	&	-0.16712	&	0.03097	&	0.04288	&	0.00986	&	0.00944	&	0.00167	&	0.00178	\\
5\%	&	0.31835	&	0.29689	&	-1.54251	&	0.56085	&	0.16997	&	-0.16719	&	0.07572	&	0.11523	&	0.03322	&	0.03194	&	0.00287	&	0.00287	\\
10\%	&	0.30473	&	0.30461	&	-1.54110	&	0.55925	&	0.16999	&	-0.16724	&	0.08781	&	0.13483	&	0.04076	&	0.03922	&	0.00397	&	0.00399	\\
20\%	&	0.26963	&	0.33352	&	-1.53119	&	0.54916	&	0.16989	&	-0.16716	&	0.14120	&	0.22378	&	0.08596	&	0.08313	&	0.00596	&	0.00598	\\
\hline
Loss of					&  \multicolumn{6}{c|}{Skewness(*)} &  \multicolumn{6}{c}{Kurtosis}		\\
\cline{2-13}	
samples	&$g_1$&$g_2$&$g_3$&$g_4$&$c_1$&$c_2$&$g_1$&$g_2$&$g_3$&$g_4$&$c_1$&$c_2$\\
\hline\hline 
0\%	&	-1.21214	&	1.23278	&	1.75298	&	-1.73202	&	-0.64331	&	0.79458	&	0.18846	&	0.19433	&	0.21259	&	0.21218	&	0.15119	&	0.16472	\\
5\%	&	-2.34607	&	1.64875	&	1.35284	&	-1.41346	&	-0.42288	&	0.36037	&	0.08094	&	0.10527	&	0.09412	&	0.09802	&	0.02540	&	0.03118	\\
10\%	&	-2.52938	&	1.97711	&	1.18018	&	-1.26045	&	-0.23379	&	0.13377	&	0.16833	&	0.17123	&	0.25041	&	0.24811	&	0.24361	&	0.23429	\\
20\%	&	-3.24122	&	1.75186	&	1.68335	&	-1.71055	&	-0.40055	&	0.37927	&	0.21292	&	0.21127	&	0.25054	&	0.24932	&	0.23883	&	0.24441	\\
\hline
\multicolumn{13}{l}{(*) Computed in accordance with the Pearson's $2^{nd}$ Coefficient of Skewness.}
\end{tabular}
}
\end{table*} 
\begin{figure*}[!ht]
\begin{center}
\begin{tabular}{cc}
\subfigure[Distribution of $|E_g|$]{
\includegraphics[trim=0cm 0cm 0cm 0.2cm, clip=true,width=.48\linewidth]{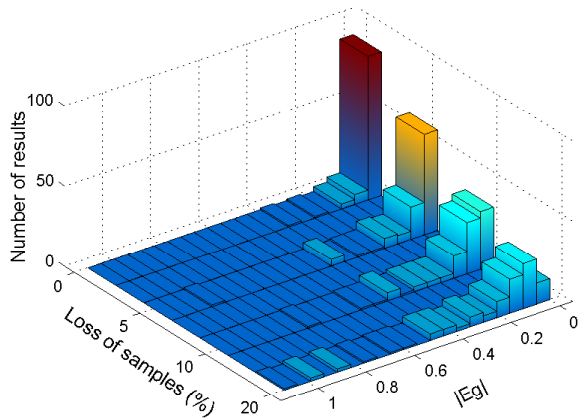}
\label{fig:hist_g1}
}&
\subfigure[Distribution of $|E_c|$]{
\includegraphics[trim=0cm 0cm 0cm 0.2cm, clip=true,width=.48\linewidth]{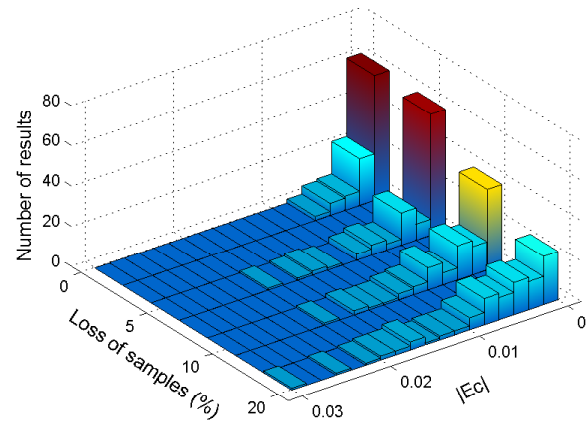}
\label{fig:hist_g2}
}\\
\end{tabular}
\caption{Histograms of $|E_g|$ and $|E_c|$ in face of different rates of sample loss}
\label{fig:hist}
\end{center}
\end{figure*}
\begin{figure*}[!ht]
\begin{center}
\begin{tabular}{cc}
\subfigure[Attack based on the data obtained without loss of samples]{
\includegraphics[trim=0cm 0cm 0cm 0.3cm, clip=true,width=.45\linewidth]{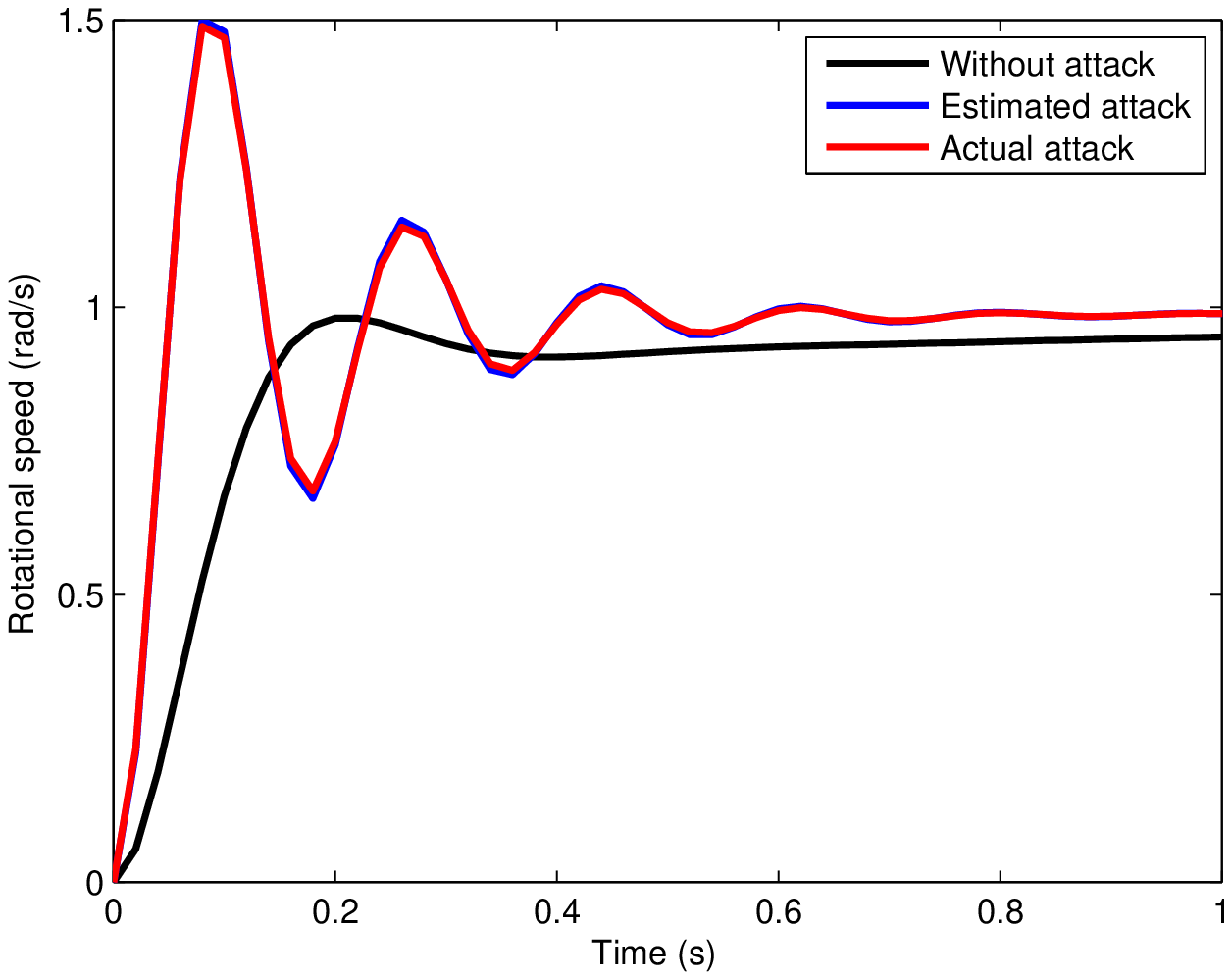}
\label{fig:hist_g1}
}&
\subfigure[Attack based on the data obtained with 20\% of sample loss]{
\includegraphics[trim=0cm 0cm 0cm 0.3cm, clip=true,width=.45\linewidth]{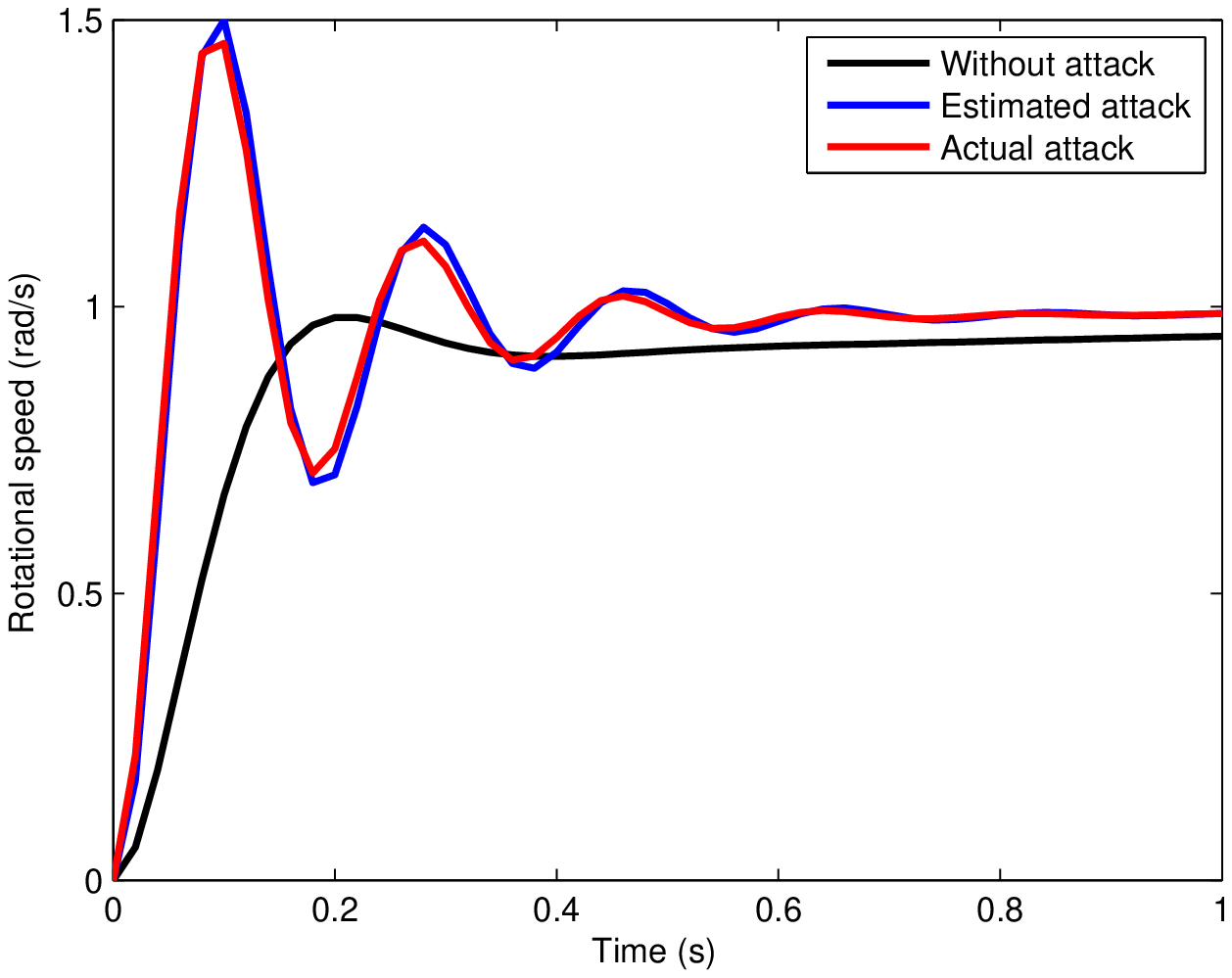}
\label{fig:hist_g2}
}\\
\end{tabular}
\vspace{-0.2cm}
\caption{Response of the system to attacks planned to cause an overshoot of 50\% in the rotational speed of the motor.}
\label{fig:ataques_tempo}
\end{center}
\end{figure*}

According with Table~\ref{tab:estatisticas} the distributions of $g_1$, $g_2$, $g_3$ and $g_4$ have a high skewness, while the distributions of $c_1$ and $c_2$ have a moderate skewness. Regarding the kurtosis, the distributions of all coefficients of $G(z)$ and $C(z)$ are leptokurtic. However, analyzing Table \ref{tab:estatisticas}, it is not possible to state a clear general pattern of an increasing/decreasing skewness or kurtosis, in face of the growth of sample loss.

In Figure \ref{fig:mean_ic}, it is possible to verify that, in all cases, the ICs tend to grow with the increase of the sample loss. The same thing occur with the standard deviations shown in Table \ref{tab:estatisticas}.
Regarding to the coefficients of $G(z)$, Figure \ref{fig:mean_ic} shows that the difference between the mean and the actual value of $g_1$, $g_2$, $g_3$ and $g_4$ also tends to raise with the increase of sample loss. It is worth mentioning that the performance of the algorithm when computing $g_3$ and $g_4$ is better then when computing $g_1$ and $g_2$, regarding the means and their CIs. This behavior results from the higher sensitivity that the output of $G(z)$ has to the variation of its poles than to the variations of its zeros. It means that, in this problem, $f_j$ grows faster for errors in $g_3$ and $g_4$ than for errors in $g_1$ and $g_2$, making the BSA population converge more accurately in dimensions $g_3$ and $g_4$.

In Figure \ref{fig:mean_ic} it is also possible to note that the accuracy obtained with the coefficients of $C(z)$ is better than the accuracy of the coefficients of $G(z)$, for all rates of sample loss. The means of $c_1$ and $c_2$ are closer to their actual values, with a smaller CI. In fact, the optimization process is more effective when computing the coefficients of $C(z)$ due to its smaller search space, that has only two dimensions instead of the four dimensions of the $G(z)$ problem.

As an additional metric to evaluate the performance of the algorithm, it is computed $|E_g|= |\mathcal{G}_r-\mathcal{G}_e|$ and $|E_c|= |\mathcal{C}_r-\mathcal{C}_e|$, that synthesize the error of the estimated coefficients of $G(z)$ and $C(z)$, respectively, for each solution found. 
$\mathcal{G}_r$ and $\mathcal{G}_e$ are vectors that contain the actual and the estimated coefficients of $G(z)$, respectively. Similarly, $\mathcal{C}_r$ and $\mathcal{C}_e$ are vectors that contain the actual and the estimated coefficients of $C(z)$, respectively. 
The histograms of $|E_g|$ and $|E_c|$ are presented in Figure \ref{fig:hist}, considering the mentioned rates of sample loss. The histograms graphically show that $|E_g|$ and $|E_c|$, which correspond to the modulus of the error of the estimated coefficients of $G(z)$ and $C(z)$, respectively, tend to present higher values as the loss of samples grows. It can also be confirmed by the increase of the standard deviation of the coefficients of $G(z)$ and $C(z)$ presented in Table \ref{tab:estatisticas}. However, according with Figure \ref{fig:hist}, the mode of this errors remain close to zero for all considered rates of sample loss.
\subsection{Results of the Service Degradation Attacks}\label{subsec:res_SD}

In this section, there are presented the results obtained through simulations of SD-Controlled Data Injection attacks, performed by a MitM acting in the control link of the NCS, as shown in Figure (\ref{fig:MitM}). The attacks were simulated in MATLAB, aiming to evaluate their accuracy when planned based on the results provided in Section \ref{subsec:res_ident}, obtained by the System Identification attack. There were performed two sets of attack. The first one, aims to cause an  {\it overshoot} of 50\% in the rotational speed of the motor. The second one, aims to cause a stationary error of $-10\%$ in the rotational speed of the motor when it is on the steady state.

In the attack aiming the overshoot, the function executed by the attacker is $M(z)=\mathcal{K}_o$. Performing a root locus analysis considering the obtained models, the attacker adjusts $\mathcal{K}_o$ to make the system underdamped, with a peak of rotational speed 50\% higher than its steady state speed. The values of $\mathcal{K}_o$ are adjusted considering the average of the coefficients estimated in Section \ref{subsec:stealthiness}. Table \ref{tab:valores_de_k} shows the values of $\mathcal{K}_o$, estimated considering different rates of sample loss during the System Identification attack, as well as the overshoots obtained with the respective $\mathcal{K}_o$ in the real model. 
In Figure  \ref{fig:ataques_tempo} it is possible to compare the response of the system without attack, with the response of the system with an attack aiming the overshoot of $50\%$. It is also possible to note that the attack to the actual model presents, in the time domain, a response quite similar to the attack estimated with the model obtained by the System Identification attack. This can be verified not only in the case where the system is identified with 0\% of sample loss, but also in the worst considered case, {\it i.e.} with 20\% of sample loss. It is worth mentioning that all responses presented in Figure~\ref{fig:ataques_tempo} converge to 1 rad/s.

In the attack where objective is to cause a stationary error of $-10\%$  on the rotational speed of the motor, the attacker executes (\ref{equ:M_z_ess}):
\vspace{-0.2cm}
\begin{equation}
M(z)= \frac{\mathcal{K}_{Ess}(z-1)}{z-0.94},
\label{equ:M_z_ess}
\end{equation}
wherein $\mathcal{K}_{Ess}$ is adjusted based on the data obtained with the System Identification attack. The pole of $M(z)$ is added aiming to allow a stationary error in the system. The zero of $M(z)$ is intended to format the root locus in order to guarantee the existence of a stable $\mathcal{K}_{Ess}$ that leads the system to a stationary error of $-10\%$. Table~\ref{tab:valores_de_k} shows the $\mathcal{K}_{Ess}$ resultant from different rates of sample loss during the System Identification attack, as well as the stationary errors obtained with the respective $\mathcal{K}_{Ess}$ in the real model.

According with the data presented in Table \ref{tab:valores_de_k}, it is possible to state that the SD-Controlled Data Injection attack, designed based on the data gathered by the System Identification attack, is capable to modify, in an accurate way, the response of the physical system, considering all the evaluated rates of sample loss. In the worst case, {\it i.e.} with 20\% of sample loss, it is obtained an overshoot of $45.94 \%$ and a stationary error of $-9.8\%$, quite close to the desired values of $50\%$ and $-10\%$, respectively. Such accuracy allows the attacker to keep his offensive under control, leading the system to a behavior that is predefined as physically covert and capable to degrade the service performed by the plant under attack.
\begin{table}[H]
\centering
\caption{Values of $\mathcal{K}_o$, $\mathcal{K}_{Ess}$ and the results obtained with the attacks}
\label{tab:valores_de_k}
\scriptsize{
\begin{tabular}{l|c|c|c|c}
\cline{2-5}	
			&  \multicolumn{4}{c}{Sample loss in the System Identification attack} \\
\cline{2-5}	
								&0 \%		&5 \%		&10	\%	&20 \%	\\
\hline\hline 
$\mathcal{K}_o$						& 4.0451	& 4.0745	& 4.0828	& 3.796 \\
Overshoot in the real model 			& 48.90 \%		& 49.43 \%		& 49.57 \%		& 45.94 \%	\\
\hline
$\mathcal{K}_{Ess}$						& 5.7471	& 5.7803	& 5.8140	& 5.8823 \\
Stationary error in the real model				& $-10\%$	& $-10\%$	& $-9.9\%$	& $-9.8\%$\\
\hline
\end{tabular}
}
\end{table} 

\section{Conclusions}\label{Conclusion}

This work proposes a physically covert attack for service degradation, which the performance depends on the knowledge about the model of the plant under attack and its controller. To obtain such knowledge, it is proposed a System Identification attack, based on the BSA algorithm. 
The effectiveness of the System Identification attack is demonstrated and its performance is statistically analyzed in face of different rates of sample loss. 
The results achieved by the physically covert attacks for service degradation, designed based on the data gathered by the System Identification attack, demonstrate the high degree of accuracy that may be achieved with the joint operation of the two attacks. In the worst case, {\it i.e.} with $20\%$ of sample loss during the System Identification attack, the attacker attained an overshoot of $45.94 \%$ and a stationary error of $-9.8\%$, quite close to the desired values of $50\%$ and $-10\%$, respectively. In both physically covert interventions, the accuracy of the attacks ensure that they will not evolve to unwanted behaviors, physically perceivable.
As future work, it is encouraged the research of techniques capable to avoid, or complicate, physically convert attacks planned with the data obtained by System Identification attacks. In this sense, we plan to further investigate countermeasures capable to make it difficult to obtain information about cyber-physical control systems, which is essential for planning covert and controlled attacks.




\bibliographystyle{IEEEtran}
\bibliography{IEEEexample}

\begin{thebibliography}{10}
\providecommand{\url}[1]{#1}
\csname url@samestyle\endcsname
\providecommand{\newblock}{\relax}
\providecommand{\bibinfo}[2]{#2}
\providecommand{\BIBentrySTDinterwordspacing}{\spaceskip=0pt\relax}
\providecommand{\BIBentryALTinterwordstretchfactor}{4}
\providecommand{\BIBentryALTinterwordspacing}{\spaceskip=\fontdimen2\font plus
\BIBentryALTinterwordstretchfactor\fontdimen3\font minus
  \fontdimen4\font\relax}
\providecommand{\BIBforeignlanguage}[2]{{%
\expandafter\ifx\csname l@#1\endcsname\relax
\typeout{** WARNING: IEEEtran.bst: No hyphenation pattern has been}%
\typeout{** loaded for the language `#1'. Using the pattern for}%
\typeout{** the default language instead.}%
\else
\language=\csname l@#1\endcsname
\fi
#2}}
\providecommand{\BIBdecl}{\relax}
\BIBdecl

\bibitem{tipsuwan2003implementation}
Y.~Tipsuwan, M.-Y. Chow, and R.~Vanijjirattikhan, ``An implementation of a
  networked pi controller over ip network,'' in \emph{Industrial Electronics
  Society, 2003. IECON'03. The 29th Annual Conference of the IEEE},
  vol.~3.\hskip 1em plus 0.5em minus 0.4em\relax IEEE, 2003, pp. 2805--2810.
  


\bibitem{gupta2010networked}
R.~A. Gupta and M.-Y. Chow, ``Networked control system: overview and research
  trends,'' \emph{Industrial Electronics, IEEE Transactions on}, vol.~57,
  no.~7, pp. 2527--2535, 2010.

\bibitem{zhang2013security}
L.~Zhang, L.~Xie, W.~Li, and Z.~Wang, ``Security solutions for networked
  control systems based on des algorithm and improved grey prediction model,''
  \emph{International Journal of Computer Network and Information Security},
  vol.~6, no.~1, p.~78, 2013.

\bibitem{farooqui2014cyber}
A.~A. Farooqui, S.~S.~H. Zaidi, A.~Y. Memon, and S.~Qazi, ``Cyber security
  backdrop: A scada testbed,'' in \emph{Computing, Communications and IT
  Applications Conference (ComComAp), 2014 IEEE}.\hskip 1em plus 0.5em minus
  0.4em\relax IEEE, 2014, pp. 98--103.

\bibitem{chow2001network}
M.-Y. Chow and Y.~Tipsuwan, ``Network-based control systems: a tutorial,'' in
  \emph{Industrial Electronics Society, 2001. IECON'01. The 27th Annual
  Conference of the IEEE}, vol.~3.\hskip 1em plus 0.5em minus 0.4em\relax IEEE,
  2001, pp. 1593--1602.

\bibitem{long2005denial}
M.~Long, C.-H. Wu, and J.~Y. Hung, ``Denial of service attacks on network-based
  control systems: impact and mitigation,'' \emph{Industrial Informatics, IEEE
  Transactions on}, vol.~1, no.~2, pp. 85--96, 2005.

\bibitem{langner2011stuxnet}
R.~Langner, ``Stuxnet: Dissecting a cyberwarfare weapon,'' \emph{Security \&
  Privacy, IEEE}, vol.~9, no.~3, pp. 49--51, 2011.

\bibitem{civicioglu2013backtracking}
P.~Civicioglu, ``Backtracking search optimization algorithm for numerical
  optimization problems,'' \emph{Applied Mathematics and Computation}, vol.
  219, no.~15, pp. 8121--8144, 2013.

\bibitem{snoeren2002single}
A.~C. Snoeren, C.~Partridge, L.~A. Sanchez, C.~E. Jones, F.~Tchakountio,
  B.~Schwartz, S.~T. Kent, and W.~T. Strayer, ``Single-packet ip traceback,''
  \emph{IEEE/ACM Transactions on Networking (ToN)}, vol.~10, no.~6, pp.
  721--734, 2002.

\bibitem{teixeira2015secure}
A.~Teixeira, I.~Shames, H.~Sandberg, and K.~H. Johansson, ``A secure control
  framework for resource-limited adversaries,'' \emph{Automatica}, vol.~51, pp.
  135--148, 2015.

\bibitem{Smth2011}
R.~Smith, ``{A} decoupled feedback structure for covertly appropriating
  networked control systems,'' in \emph{Proceedings of the 18th IFAC World
  Congress 2011}, vol.~18, no.~1.\hskip 1em plus 0.5em minus 0.4em\relax
  IFAC-PapersOnLine, 2011.

\bibitem{smith2015covert}
R.~S. Smith, ``Covert misappropriation of networked control systems: Presenting
  a feedback structure,'' \emph{Control Systems, IEEE}, vol.~35, no.~1, pp.
  82--92, 2015.

\bibitem{hussain2003framework}
A.~Hussain, J.~Heidemann, and C.~Papadopoulos, ``A framework for classifying
  denial of service attacks,'' in \emph{Proceedings of the 2003 conference on
  Applications, technologies, architectures, and protocols for computer
  communications}.\hskip 1em plus 0.5em minus 0.4em\relax ACM, 2003, pp.
  99--110.

\bibitem{ramos2011cyber}
C.~Ramos, Z.~Vale, and L.~Faria, ``Cyber-physical intelligence in the context
  of power systems,'' in \emph{Future Generation Information Technology}.\hskip
  1em plus 0.5em minus 0.4em\relax Springer, 2011, pp. 19--29.

\bibitem{khatri2015taxonomy}
S.~Khatri, P.~Sharma, P.~Chaudhary, and A.~Bijalwan, ``A taxonomy of physical
  layer attacks in manet,'' \emph{International Journal of Computer
  Applications}, vol. 117, no.~22, 2015.

\bibitem{zou2016intercept}
Y.~Zou and G.~Wang, ``Intercept behavior analysis of industrial wireless sensor
  networks in the presence of eavesdropping attack,'' \emph{IEEE Transactions
  on Industrial Informatics}, vol.~12, no.~2, pp. 780--787, 2016.

\bibitem{de2016distributed}
A.~O. de~S{\'a}, N.~Nedjah, and L.~de~Macedo~Mourelle, ``Distributed efficient
  localization in swarm robotic systems using swarm intelligence algorithms,''
  \emph{Neurocomputing}, vol. 172, pp. 322--336, 2016.

\bibitem{stallings2006cryptography}
W.~Stallings, \emph{Cryptography and network security: principles and
  practices}.\hskip 1em plus 0.5em minus 0.4em\relax Pearson Education India,
  2006.

\bibitem{el1989variable}
M.~El-Sharkawi and C.~Huang, ``Variable structure tracking of dc motor for high
  performance applications,'' \emph{Energy Conversion, IEEE Transactions on},
  vol.~4, no.~4, pp. 643--650, 1989.

\bibitem{tran2007robust}
T.~Tran, Q.~P. Ha, and H.~T. Nguyen, ``Robust non-overshoot time responses
  using cascade sliding mode-pid control,'' \emph{Journal of Advanced
  Computational Intelligence and Intelligent Informatics}, 2007.

\bibitem{hwang2008study}
H.~Hwang, G.~Jung, K.~Sohn, and S.~Park, ``A study on mitm (man in the middle)
  vulnerability in wireless network using 802.1 x and eap,'' in
  \emph{Information Science and Security, 2008. ICISS. International Conference
  on}.\hskip 1em plus 0.5em minus 0.4em\relax IEEE, 2008, pp. 164--170.

\end{thebibliography}

%

%






\end{document}